\documentclass{iopconfser}

\usepackage{graphicx}
\usepackage{newtxtext}
\usepackage{newtxmath}
\usepackage{hyperref}
\usepackage{enumerate}
\usepackage{fancyhdr}
\begin{document}
\title{Analytical Solutions for Turbulent Channel Flow Using Alexeev and Navier-Stokes Hydrodynamic Equations: Comparison with Experiments}

% MDPI internal command: Title for citation in the left column
%\TitleCitation{Analytical Solutions for Turbulent Channel Flow Using Navier-Stokes and Alexeev Hydrodynamic Equations and Their Comparison}

% Author Orchid ID: enter ID or remove command
%\newcommand{\orcidauthorA}{0000-0002-5107-8532} % Add \orcidA{} behind the author's 
% 
% Authors, for the paper (add full first names)
\author{Alex Fedoseyev}%  \orcidA{0000-0002-5107-8532}} 
\affil{Ultra Quantum Inc., Huntsville, Alabama, USA}
\email{af@ultraquantum.com}
%
% Author citation:  
%\AuthorCitation{Fedoseyev, A.}

% Affiliations / Addresses (Add [1] after \address if there is only one affiliation.)
%\address{%
% \quad Ultra Quantum Inc, Huntsville, Alabama, USA; af@ultraquantum.com}

% Contact information of the corresponding author
%\corres{Correspondence: af@ultraquantum.com}
%
%\title{Analytical Solutions for Turbulent Channel Flow Using Navier-Stokes and Alexeev Hydrodynamic Equations}

%\author{Alex Fedoseyev}
%  \email[Corresponding author: ]{af@ultraquantum.com}
%\affiliation{Ultra Quantum Inc., Huntsville, Alabama, USA
%}

\begin{abstract}
Understanding turbulent boundary layer flows is important for many application areas. Enhanced theoretical models may provide deeper insights into the fundamental mechanisms of turbulence that elude current models; therefore, the search for improved kinetic equations and their respective hydrodynamic equations continues. In this work, we consider the Generalized Boltzmann Equation (GBE), proposed by Alexeev (1994). The GBE accounts for finite particle size and the variation of the distribution function over timescales of the order of the collision time. The Alexeev hydrodynamic equations are derived from the GBE. 
In this work, the Alexeev hydrodynamic equations (AHE) and Navier-Stokes (NS) equations are solved analytically for turbulent channel flow under the assumption that stationary solutions yield the mean flow velocity. The analytical solutions of the AHE are validated by numerical solutions and compared with the NS solutions and experimental data  for turbulent channel flow from multiple sources, spanning Reynolds numbers from 3,000 to 35,000,000.
Solutions of the AHE demonstrate significantly better agreement with experimental data than those obtained from the NS equations. The analytical solution revealed a new similarity parameter: the boundary layer thickness scale, which coincides with the Kolmogorov microscale observed in experiments. The mechanisms for turbulence generation and control are discussed.

% Keywords
%\keyword{

\end{abstract}

%%%%%%%%%%%%%%%%%%%%%%%%%%%%%%%%%%%%%%%%%%
%\begin{document}

%%%%%%%%%%%%%%%%%%%%%%%%%%%%%%%%%%%%%%%%%%
%\setcounter{section}{0} %% Remove this when starting to work on the template.

%%%%%%%%%%%%%%%%%%%%%%%%%%%%%%%%%%%%%%%%%%%%%%%%%%%%%%%%%%%%%%%%%%%%%%%%%%%%%%%%%%
%
%    Section Introduction
%
%%%%%%%%%%%%%%%%%%%%%%%%%%%%%%%%%%%%%%%%%%%%%%%%%%%%%%%%%%%%%%%%%%%%%%%%%%%%%%%%%%
\section{Introduction\label{sec:intro}}

Understanding turbulent boundary layer flows is important for many application areas. Experimental investigations of turbulent channel flow provide critical data for improving theoretical and numerical models and offer insights into the fundamental mechanisms of turbulence that may not be captured by existing approaches. Enhanced theoretical models may provide deeper insights into these fundamental mechanisms; therefore, the search for improved kinetic equations, their respective hydrodynamic equations, and accurate turbulence models continues.

The proposed approach for turbulent flow solutions uses a novel theoretical model: the Alexeev hydrodynamic equations (AHE) obtained by Alexeev in 1994 \cite{Alexeev_1994}. The AHE were derived from the Generalized Boltzmann Equation (GBE) \cite{Alexeev_1994}. The GBE differs from the Boltzmann Equation (BE) by accounting for finite particle size and the variation of the distribution function over timescales of the order of the collision time. Consequently, the GBE includes a second material derivative term with a timescale coefficient.

We begin with a review of the Generalized Boltzmann Equation, its derivation and properties, and then present the Alexeev hydrodynamic equations, including the full system for incompressible flow and several simplified models. The validation of the AHE is referenced to several different authors.
The derivation of the analytical solution of the AHE for turbulent channel flow is provided, and the solution is compared to the numerical solution of the AHE for validation.
The results are compared to the Navier-Stokes (NS) solutions and to high-quality experimental data from multiple sources.
The conclusion summarizes the results.

%%%%%%%%%%%%%%%%%%%%%%%%%%%%%%%%%%%%%%%%%%%%%%%%%%%%%%%%%%%%%%%%%%%%%%%%%%%%%%%%%%
%
%    Section Generalized Boltzmann Equation
%
%%%%%%%%%%%%%%%%%%%%%%%%%%%%%%%%%%%%%%%%%%%%%%%%%%%%%%%%%%%%%%%%%%%%%%%%%%%%%%%%%%
\section{Generalized Boltzmann Equation \label{sec:GBE}}

The standard Boltzmann transport equation accounts for changes in the distribution function  $f({\bf r, v,}  t)$ on the hydrodynamic timescale and the mean time between collisions of infinitesimal particles, which are treated as material points. Introducing a third timescale associated with the finite dimensions of interacting particles leads to an additional term in the equation: a second material derivative of the distribution function. This results in a generalized form known as the Generalized Boltzmann Equation \cite{Alexeev_1994}:

\begin{equation}
\label{eq:GBE}%%\begin{align}
\frac{ Df}{Dt} - \frac{D}{Dt}\left(\tau\frac{ Df}{Dt}\right)= J^{B},
%%\end{align}
\end{equation}

\noindent
where $f=f({\bf r, v,}  t)$ is particle velocity distribution function, and operator
\begin{equation}
\frac{D}{Dt} = \frac{\partial}{\partial t} + \bf v \cdot \frac{\partial}{\partial \bf r} + \bf F \cdot \frac{\partial}{\partial \bf v}
\end{equation}
 represents material derivative in space, velocity space and time, and $J^B$ is the collision integral, the same as in Boltzmann equation (\cite{Alexeev_2004}, p.11), where {\bf v} and {\bf r} are  the velocity and the position of the particle, respectively, $t$ is the time, and {\bf F} is the force acting on the particle. A timescale multiplier $\tau$ is a material property. 
 
The GBE was obtained from first principles, starting from the Liouville equation for the multi-particle distribution function $f_N$ of a system of $N$ interacting particles by introducing a small parameter $\varepsilon =n v_b$, where $n$ is the number particle density, and $v_b$ is the interaction volume. Then, starting from the hierarchy of Bogolyubov kinetic equations (chain), Alexeev broke the Bogolyubov chain at the $r_b$-scale,where $r_b = v_b^{1/3}$, finding the exact representation of the term \cite{Alexeev_2004}(p.52), and obtained a one-particle representation function $f({\bf r, v,}  t)$ resulting in the GBE equation  ($\ref{eq:GBE}$). The timescale $\tau$ is introduced as:
 
 \begin{equation}
\label{eq:taub} 
\tau = \frac{ \varepsilon}{[\partial \varepsilon/\partial t]_{t=0}},
\end{equation}
where $\varepsilon$ is the number of particles of all kind, found themself within the interaction volume of a particle numbered as $\it 1$ at the instant of time $t$, and the derivative $[\partial \varepsilon/\partial t]_{t=0}$ is taken at the moment $t=0$, when the volume $v_b$ is empty (no particles).

During the derivation of the GBE, the sign of the $\tau$ multiplier must be chosen, and the negative sign was selected to satisfy the Boltzmann H-theorem. The GBE is thermodynamically consistent and satisfies the conditions for the Boltzmann H-theorem \cite{Alexeev_2004}.

Alexeev assumed that the parameter $\tau$ defined in this way is the mean time between collisions \cite{Alexeev_2004}(p.52). This is not correct. It was found that the parameter $\tau$ is of much larger value, and is a material property, that, in the case of the hydrodynamic equations, gives rise to the length scale $\delta= \sqrt{\tau \nu}$, where $\nu$ is the kinematic viscosity. We will show that $\delta$ is a boundary layer thickness scale and that $\delta$ coincides with the Kolmogorov microscale observed in experiments. This $\delta$ has a value of approximately 0.6 mm for distilled water and approximately the same for air over a wide pressure range (to hundreds of atmospheres). 

The spacial range of validity of the GBE extends from hydrodynamic scale $L$ down to the mean free path between collisions $\lambda$, and further down to the particle interaction scale $r_b$ \cite{Alexeev_2004}. The Alexeev hydrodynamic equations have the same range of validity as the GBE, which has been confirmed by simulations for rarefied hypersonic flows \cite{Fedoseyev_2020, Fedoseyev_2021, Fedoseyev_2022}.

%%%%%%%%%%%%%%%%%%%%%%%%%%%%%%%%%%%%%%%%%%%%%%%%%%%%%%%%%%%%%%%%%%%%%%%%%%%%%%%%%%
%
%    Section Alexeev Hydrodynamic Equations
%
%%%%%%%%%%%%%%%%%%%%%%%%%%%%%%%%%%%%%%%%%%%%%%%%%%%%%%%%%%%%%%%%%%%%%%%%%%%%%%%%%%
\section{Hydrodynamic Equations \label{sec:AHE}}
%%%%%%%%%%%%%%%%%%%%%%%%%%%%%%%%%%%%%%%%%%%%%%%
%
%    Equations: Alexeev Hydrodynamic Equations
%
%%%%%%%%%%%%%%%%%%%%%%%%%%%%%%%%%%%%%%%%%%%%%%%
\subsection{Alexeev Hydrodynamic Equations}
The Alexeev Hydrodynamic Equations (AHE) were derived from the Generalized Boltzmann Equation by multiplying the latter by the standard collision invariants (mass, momentum, energy) and integrating the result over the velocity space. For incompressible flow and thermal convection, the AHE were first presented by Fedoseyev and Alexeev (1998) in \cite{Fedoseyev_1998a, Fedoseyev_1998b}. The main results from these papers were summarized in a book by Alexeev (2004) \cite{Alexeev_2004} (p. 253). The full AHE were provided in \cite{Fedoseyev_1998a} and \cite{Fedoseyev_2012}, and they have the following form:

%%%%%%%%%%%%%%%%%%%%%%%%%%%%%%%%%%%%%%%%%%%%%%%
%
%    Equations: momentum and continuity
%
%%%%%%%%%%%%%%%%%%%%%%%%%%%%%%%%%%%%%%%%%%%%%%%
%\vspace{-0.25cm}
\begin{equation}
\frac{\partial {\bf V}} {\partial {\rm t}} + ({\bf V}\cdot \nabla) {{\bf V}}
- \frac{1}{Re}\nabla^2 {\bf V} + \nabla {\rm p} - {\bf F} = 
{\tau} \left\{ \frac{\partial^2 {\bf V}} {\partial {\rm t^2}} + 2 \frac{\partial}{\partial t}(\nabla {\rm p}) + 
\nabla^2 ({\rm p} {\bf V}) + \nabla(\nabla \cdot ({\rm p} {\bf V)}) 
\right\}
\label{eq:momeq}
\end{equation}
\noindent
while continuity equation is
%\vspace{-0.5cm}
\begin{equation}
\nabla \cdot {\bf V} = 
{\tau} \left\{
2 \frac{\partial}{\partial t}(\nabla \cdot {\bf V})
+ \nabla \cdot ({\bf V} \nabla){\bf V}
+\nabla^2 {\rm p} -\nabla \cdot {\bf F}
\right\}
\label{eq:newcont}
\end{equation}

\noindent
where ${\bf V}$ and $p$ are nondimensional velocity and pressure, ${Re = V_0 L_0/\nu}$ - the Reynolds number, $V_0$ - velocity scale, $L_0$ - 
hydrodynamic length scale, $\nu$ - kinematic viscosity, ${\bf F}$ is
a body force and a nondimensional $\tau$ is
\begin{equation}
\tau = \tau_0 L_0^{-1}V_0 = \tau_0 \nu /L_0^2 Re = \delta^2 Re,
\label{eq:tau}
\end{equation}
where $\tau_0 $ is the dimensional value of $\tau$.
The parameter  $\tau$ is assumed to be constant. Note that the AHE have two similarity parameters: the Reynolds number $Re$ and the boundary layer thickness scale $\delta$,
\begin{equation}
\delta = \sqrt{\tau_0 \nu} /L_0 .
\label{eq:delta}
\end{equation}
Additional boundary conditions on the walls require that the fluctuations be zero:
The boundary condition for pressure on the walls is
%\vspace{-0.25cm}
\begin{equation} \label{eq:p}
 {(\nabla \rm p - {\bf F})\cdot {\bf n}= 0} ,
\end{equation}
%\vspace{-0.25cm}
where ${\bf n}$ is a wall normal. These equations have been used to obtain a transient analytical solution for the transverse velocity component \cite{Fedoseyev_2024c}.

%%%%%%%%%%%%%%%%%%%%%%%%%%%%%%%%%%%%%%%%%%%%%%%
%
%    Equations: Alexeev Hydrodynamic Equations
%
%%%%%%%%%%%%%%%%%%%%%%%%%%%%%%%%%%%%%%%%%%%%%%%
\subsection{Simplified Alexeev Hydrodynamic Equations}
A simplified model of the AHE for incompressible fluid was first presented in \cite{Fedoseyev_2001a, Fedoseyev_2010}. This model has been successfully applied to a number of high Reynolds number flows and thermal convection problems, as well as to magneto-hydrodynamics \cite{Fedoseyev_2001b, Ananiev_2004, Ananiev_2005, Ananiev_2005b, Volkov_2011}. We use this simplified model for stationary turbulent channel flow:
\begin{equation}
\it \frac{\partial \bf V} {\partial \rm t} + ({\bf V}\cdot \nabla) {\bf V}
- \frac{1}{Re}\nabla^2 \bf V + \nabla \rm p  = 0,
\label{eq:momeq}
\end{equation}
\begin{equation}
\nabla \cdot {\bf V} = {\tau} \nabla^2 \rm p ,
\label{eq:cont}
\end{equation}
\noindent
where ${\bf V}$ and $p$ are nondimensional velocity and pressure respectively, ${Re=U_0 L_0/\nu}$ - the Reynolds number, $U_0$ - velocity scale, $L_0$ - 
hydrodynamic length scale, $\nu$ - kinematic viscosity, and nondimensional timescale ${\tau = \tau_0 L_0^{-1}U_0}$, where $\tau_0$ is the dimensional timescale, a material property. The boundary condition for pressure on the walls is $(\nabla \rm p \cdot {\bf n}) = 0$. 

The timescale $\tau$ can be rewritten as 
\begin{equation}\label{eq:tau}
\tau = \delta^2 Re,
\end{equation}
 where  
\begin{equation}
\delta = \frac{\sqrt{\tau_0 \nu}}{L_0},
\label{eq:delta}
\end{equation}
another similarity parameter together with the Reynolds number.
Equations (\ref{eq:momeq}) and (\ref{eq:cont})  reduce to the Navier-Stokes equations when $\tau = 0$, and their solution becomes the solution of the Navier-Stokes equations. That has been proven recently by Amosova (2023) \cite{Amosova_2023}.
%%%%%%%%%%%%%%%%%%%%%%%%%%%%%%%%%%%%%%%%%%%%%%%%%%%%%%%%%%%%%%%%%%%%%%%%%%%%%%%%%%
%
%    Section Two-dimensional channel flow
%
%%%%%%%%%%%%%%%%%%%%%%%%%%%%%%%%%%%%%%%%%%%%%%%%%%%%%%%%%%%%%%%%%%%%%%%%%%%%%%%%%%
\section{Governing equations for two-dimensional channel flow and solutions \label{sec:2d_chan}}
 
For the case of two-dimensional horizontal channel flow with width $L_0$ , we seek stationary solutions in half of the channel (assuming solution symmetry), under the assumption that these solutions yield the mean flow velocity. In this configuration, no variables depend on $x$ (the streamwise direction), so all derivatives with respect to $x$ vanish except for the pressure gradient $p_x  = const$. The problem thus reduces to a one-dimensional case where all the variables depend only on $y$. 
%%%%%%%%%%%%%%%%%%%%%%%%%%%%%%%%%%%%%%%%%%%%%%%%%%%%%%%%%%%%%%%%%%%%%%%%%%%%%%%%%%
%
%    Subsection Navier-Stokes equations
%
%%%%%%%%%%%%%%%%%%%%%%%%%%%%%%%%%%%%%%%%%%%%%%%%%%%%%%%%%%%%%%%%%%%%%%%%%%%%%%%%%%

\subsection{Stationary Navier-Stokes equations for 2D flow in channel\label{sec:2d_ns}}

In the continuity equation of the Navier-Stokes equations, 

\begin{equation}
\label{sec:2d_cont}
\nabla \cdot {\bf V} = 0
\end{equation}

for velocity ${\bf V} = (U,V)$, the derivative $U_x=0$, and thus
\begin{equation}
\label{sec:2d_Vy}
V_y = 0.
\end{equation}

\noindent
With the boundary condition $V(0)=0$, this yields the solution $V=0$ for transverse flow velocity component. The momentum equation then becomes
\begin{equation}
\label{eq:2d_mom}
- \frac{1}{Re} U_{yy} + p_x  = 0,
\end{equation}
with the boundary condition $U(0)=0$. Assuming the velocity at the channel centerline is $U=U_0$, the solution can be readily obtained as 
\begin{equation}
\label{eq:2d_ns_sol}
U(y) = U_0\cdot4y(L-y)/L^2,
\end{equation}
\noindent
where $U_0=\frac{1}{8} Re p_x L_0^2$.  The well-known parabolic solution for the streamwise velocity was obtained, representing the laminar flow. No other solutions are possible for the stationary Navier-Stokes equations.
%%%%%%%%%%%%%%%%%%%%%%%%%%%%%%%%%%%%%%%%%%%%%%%%%%%%%%%%%%%%%%%%%%%%%%%%%%%%%%%%%%
%
%    Subsection Alexeev hydrodynamic equations
%
%%%%%%%%%%%%%%%%%%%%%%%%%%%%%%%%%%%%%%%%%%%%%%%%%%%%%%%%%%%%%%%%%%%%%%%%%%%%%%%%%%

\subsection{Stationary Alexeev hydrodynamic equations for 2D flow in channel\label{sec:2d_ahe}}

The AHE become the following: the momentum and  continuity equations are
\begin{equation}
\label{eq:2d_mom_U}
V U_y - \frac{1}{Re} U_{yy} + p_x  = 0,
\end{equation}
\begin{equation}
\label{eq:2d_mom_V}
V V_y - \frac{1}{Re} V_{yy} + p_y = 0,
\end{equation}
\noindent
\begin{equation}
\label{eq:2d_cont_ahe}
V_y = \tau p_{yy}.
\end{equation}

\noindent
with the boundary condition $V(0)=0$,  $U(0)=0$, $U(0.5)=U_0$.

%%%%%%%%%%%%%%%%%%%%%%%%%%%%%%%%%%%%%%%%%%%%%%%%%%%%%%%%%%%%%%%%%%%%%%%%%%%%%%%%%%
%
%    Subsection Analytical solution of Alexeev hydrodynamic equations
%
%%%%%%%%%%%%%%%%%%%%%%%%%%%%%%%%%%%%%%%%%%%%%%%%%%%%%%%%%%%%%%%%%%%%%%%%%%%%%%%%%%
\subsection{Approximate analytical solution of the Alexeev hydrodynamic equations \label{sec:anal_sol}}
Finding an analytical solution to equations (\ref{eq:2d_mom_U}), (\ref{eq:2d_mom_V}), and (\ref{eq:2d_cont_ahe}) is difficult. To simplify the problem, we replace these equations with two sets of equations describing the laminar and turbulent velocity components. We assume that the general solution $U(y)$ of equations (\ref{eq:2d_mom_U}), (\ref{eq:2d_mom_V}), and (\ref{eq:2d_cont_ahe}) is a superposition of the laminar and turbulent velocity solutions, $U^L$ and  $U^T$, 
\begin{equation}
\label{eq:2d_usol}
U = U_0(\gamma U^T + (1-\gamma) U^L),
\end{equation}
along with $V$
\begin{equation}
\label{eq:2d_vsol}
V = \gamma V^T,
\end{equation}
since $V^L=0$.
The AHE equations become the following:

\begin{equation}
\label{eq:2d_U}
V^T (\gamma U^T_y + (1-\gamma) U^L_y)- \frac{1}{Re} (\gamma U^T{yy} + (1-\gamma) U^L_{yy}) + p_x  = 0,
\end{equation}
\begin{equation}
\label{eq:2d_V}
V^T V^T_y - \frac{1}{Re} V^T_{yy} + p^T_y = 0,
\end{equation}
\noindent
\begin{equation}
\label{eq:2d_cont_ahe2}
V^T_y = \tau p^T_{yy}.
\end{equation}

The first set consists of one equation: the laminar velocity component equation for $U^L$:
\begin{equation}
\label{eq:2d_mom_UL}
- \frac{1}{Re} (1-\gamma)U^L_{yy} + p_x  = 0,
\end{equation}
with the boundary conditions $U^L(0)=0$, $U^L(0.5)=U_0$, and the transverse velocity component $V^L = 0$.
%\begin{equation}
%\label{sec:2d_cont_L}
%V^L  = 0.
%$end{equation}
The solution of this equation is given by (\ref{eq:2d_ns_sol}).

The second set consists of the turbulent velocity component equations -- the  momentum equations for $U^T$ and $V^T$ and the continuity equation --  and has the form:
\begin{equation}
\label{eq:2d_mom_UT}
V^T (U^T_y + \alpha U^L_y) - \frac{1}{Re} U^T_{yy}   = 0,\\
\end{equation}
\begin{equation}
\label{eq:2d_mom_VT}
V^T V^T_y - \frac{1}{Re} V^T_{yy} + p^T_y = 0,\\
\end{equation}
\begin{equation}
V^T_y = \delta^2Re \cdot  p^T_{yy},
\label{sec:2d_cont_VT}
\end{equation}
where $\tau$ was substituted from (\ref{eq:tau}) by $\delta^2Re$, 
and $\alpha = (1-\gamma)/\gamma < 1$ is the fraction of the laminar velocity component relative to the turbulent one. The boundary conditions are  $U^T(0)=0$, $V^T(0)=0$, $U^T(0.5)=U_0$.
Equations (\ref{eq:2d_mom_VT}) and (\ref{sec:2d_cont_VT}) can be combined, resulting in an independent equation for $V^T$,
\begin{equation}
\label{eq:2d_VT}
\frac{1}{Re} V^T_{yy} -V^T V^T_y -\frac{1}{\delta^2 Re} V^T = 0.
\end{equation}

We can see that the streamwise component of velocity $U^T$ in the linear equation (\ref{eq:2d_mom_UT}) depends on $V^T$, while the equation for the transverse velocity component  $V^T$ is completely independent.  Equation~(\ref{eq:2d_VT}) can be solved, and its  solution for $V^T(y)$ can be used to solve equation~(\ref{eq:2d_mom_UT}) for $U^T(y)$. 

If the second term in the equation~(\ref{eq:2d_mom_UT}) containing $U^L_y$ is neglected, since $\alpha U^L_y \ll U^T_y$ (in the boundary layer), the solution $U^T$ can be obtained by integrating equation~(\ref{eq:2d_mom_UT}), resulting in an analytical formula that expresses the solution $U^T(y)$ through integrals of $V^T(y)$,
%\vspace{-0.35cm}
\begin{equation}\label{eq:u_sol3}
U^T(y) = C_1\int_1^y  \exp{ \left(\int_1^\eta Re \cdot  V^T(\xi) d \xi \right)} d\eta + C_2,
\end{equation}
where the constants $C_1$ and $C_2$ are chosen to satisfy the boundary conditions. Alternatively, equation~(\ref{eq:2d_mom_UT}) can be solved by numerical methods, e.g. the finite-difference method.

Non-trivial approximate analytical solutions of equations (\ref{eq:2d_mom_UT}) to (\ref{sec:2d_cont_VT}) were derived in \cite{Fedoseyev_2023}, which presented the solution as  an explicit formula:
\begin{equation}\label{eq:UTsol}
U^T=U_{0}\left(1-e^{1-e^{y/\delta}}\right),
\end{equation}
\begin{equation}\label{eq:VTsol}
V^T=\frac{1}{\delta Re}\left({1-e^{y/\delta}}\right),
\end{equation}
\begin{equation}\label{eq:pTsol}
p^T_y=\frac{1}{\delta^2 Re}V^T.
\end{equation}

%%%%%%%%%%%%%%%%%%%%%%%%%%%%%%%%%%%%%%%%%%%%%%%%%%%%%%%%%%%%%%%%%%%%%%%%%%%%%%%%%%
%
%    Subsection Validation for turbulent solution
%
%%%%%%%%%%%%%%%%%%%%%%%%%%%%%%%%%%%%%%%%%%%%%%%%%%%%%%%%%%%%%%%%%%%%%%%%%%%%%%%%%%
\subsection{Validation for the turbulent solution $U^T$ \label{sec:validation}}
The approximate solution for $V^T$, equation (\ref{eq:VTsol}), was obtained in \cite{Fedoseyev_2023} from equation (\ref{eq:2d_mom_VT}) by dropping the nonlinear term $V^T V^T_y$, differentiating it with respect to $y$, and substituting the expression for $ p^T_{yy}$ into  equation (\ref{sec:2d_cont_VT}), obtaining the equation  

\begin{equation}
\label{eq:2d_VT_simple}
V^T_{yyy} -\frac{1}{\delta^2} V^T_y = 0.
\end{equation}

\noindent 
When the sign of $\delta=\sqrt{\delta^2}$ was taken as positive, the solution $V^T$ is
\begin{equation}
\label{eq:2d_VT_sol1}
V^T=\frac{1}{\delta Re}\left({1-e^{y/\delta}}\right).
\end{equation}

\noindent Another solution, when the sign of $\delta=-\sqrt{\delta^2}$ was taken as negative, was considered in \cite{Fedoseyev_2025}:

\begin{equation}
\label{eq:2d_VT_sol2}
V^T= -\frac{1}{\delta Re}\left({1-e^{-y/\delta}}\right).
\end{equation}
The linear combination of $U^T$ solutions, corresponding to the $V^T$ solutions above, gave more accurate results for $U^T$ at high Re numbers \cite{Fedoseyev_2024}. 

%\noindent 
When the $V^T$ solution (\ref{eq:2d_VT_sol1}) was substituted into the simplified equation (\ref{eq:2d_mom_UT}) with omitted term $\alpha U^L_y$, since $\alpha U^L_y \ll U^T_y$ (in the boundary layer), it was possible to solve it analytically, 
resulting in the solution  $U^T$ (\ref{eq:UTsol}). Dropping the term $V^T V^T_y$  was a significant simplification but necessary step to obtain a closed-form formula. Therefore,  validation of the  $U^T$ solution is required. 

In the following, the  numerical solution of the equations for $V^T$ and $U^T$ from the more accurate equations is presented: 

\begin{equation}
\label{eq:2d_VT2a}
\frac{1}{Re} V^T_{yy} -V^T V^T_y -\frac{1}{\delta^2 Re} V^T = 0,
\end{equation}

\begin{equation}
\label{eq:2d_mom_UT2}
V^T U^T_y - \frac{1}{Re} U^T_{yy}   = 0. 
\end{equation}
The solution was computed for the Wei (1989) experiment at Re = 14914 using a finite-difference method with N = 201 to 401 nodes (yielding indistinguishable solutions) on a non-uniform mesh refined near the boundaries. To avoid the trivial solutions $U^T$ and $V^T=0$, the deflation technique proposed by Farrell (2015) was applied \cite{Farrell_2015}.

Equation (\ref{eq:2d_VT2a}) results in a nonlinear system of finite-difference equations, which was solved by minimizing the squared residual using the Powell hybrid method (MINPACK) \cite{Powell_1964}. The same approach was used to solve equation (\ref{eq:2d_mom_UT2}).

\begin{figure}[h!]
%\vspace {-7mm}
	\begin{center}
        \includegraphics[width=0.49\linewidth]{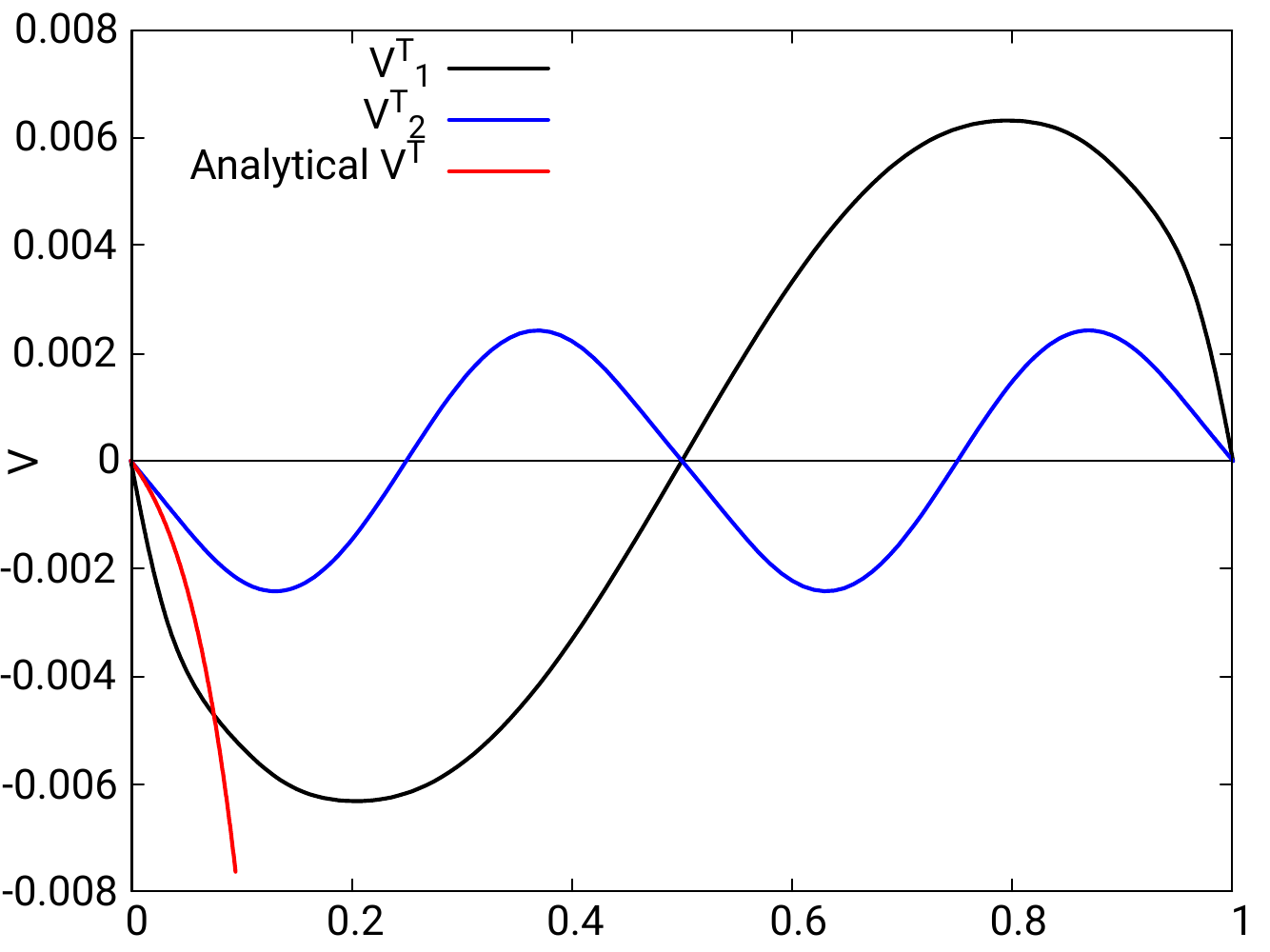}
        \includegraphics[width=0.49\linewidth]{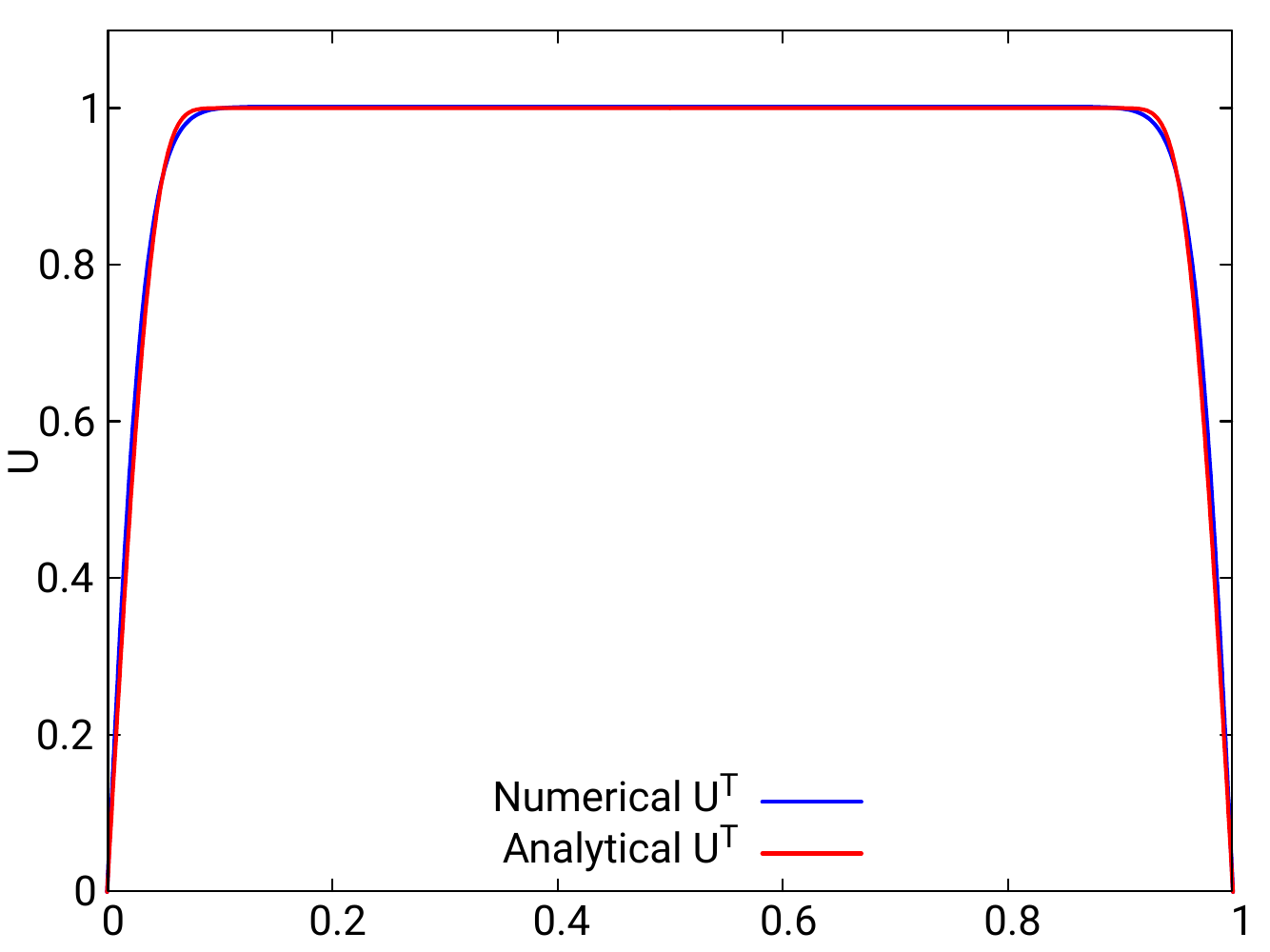}
\hspace{2.5cm} (a) \hspace{6.5cm}(b)
%\vspace {-5mm}
	\caption{\label{fig:VTUT_sol} (a) Two numerical solutions for the transverse velocity $V^T(y)$ in the channel at Re = 14914, computed using equation (\ref{eq:2d_VT2a}). The initial approximations were $sin(n\pi y)$, for $n = 2, 4$. The analytical solution  (\ref{eq:VTsol}) is also shown (red line). (b) The solution for the streamwise turbulent velocity $U^T(y)$ from equation (\ref{eq:2d_mom_UT2}) (blue line), corresponding to the $V^T_1(y)$ solution (the only non-zero numerical solution), and the analytical solution $U^T(y)$ (\ref{eq:UTsol}) (red line) are shown. }
	\end{center}
%\vspace {-6mm}
\end{figure}

It was found that equation (\ref{eq:2d_VT2a}) admits multiple solutions; two of them are shown in  Figure \ref{fig:VTUT_sol}(a). Only one of them, $V^T_1$ , results in the non-zero $U^T(y)$ solution shown by the blue line in Figure \ref{fig:VTUT_sol}(b). 
The analytical solution $U^T(y)$ (\ref{eq:UTsol}) is also shown in Figure~\ref{fig:VTUT_sol}(b) as the red line and is very close to the numerical solution, nearly indistinguishable in the figure. 

The validation provided above confirms that the approximate analytical solution $U^T$  is in excellent agreement with the numerical solution for this Reynolds number. For very high Reynolds numbers ranging from 1,000,000 to 35,000,000, there is a larger deviation between the numerical and analytical $U^T$ solutions; for such cases, an improved analytical solution is proposed \cite{Fedoseyev_2025}. 

%%%%%%%%%%%%%%%%%%%%%%%%%%%%%%%%%%%%%%%%%%%%%%%%%%%%%%%%%%%%%%%%%%%%%%%%%%%%%%%%%%
%
%    Subsection General  solution
%
%%%%%%%%%%%%%%%%%%%%%%%%%%%%%%%%%%%%%%%%%%%%%%%%%%%%%%%%%%%%%%%%%%%%%%%%%%%%%%%%%%
\subsection{General solution of Alexeev hydrodynamic equations \label{sec:general}}
The general solution $U(y)$ for turbulent flow in channel is proposed as a linear superposition of $U^L$ and $U^T$ \cite{Fedoseyev_2023},
\begin{eqnarray}\label{eq:AHEsol}
U(y)=U_{0}\left[\gamma\left(1-e^{1-e^{y/\delta}}\right)+(1-\gamma)4y(L-y)/L^{2}\right],
\end{eqnarray}
\noindent where $\gamma$ and $(1-\gamma)$, the coefficients of superposition, that are introduced to satisfy $U(0.5)= U_0$. 
The analytical solution (\ref{eq:AHEsol}) can also be used for the turbulent flow in circular pipe if $\delta \ll 1$ \cite{Fedoseyev_2025}. The parameter $\gamma$ is calculated through the principle of minimum viscous dissipation \cite{Fedoseyev_2024}, and typically $\gamma$ is close to 0.65. The dimensional value of $\delta$, $\delta_0=\sqrt{\tau_0 \nu}$ is approximately 0.6 mm for distilled water and air over a wide pressure range \cite{Fedoseyev_2025}.

%====================================================================
\begin{figure}
\begin{center}
\includegraphics[width=0.65\textwidth]{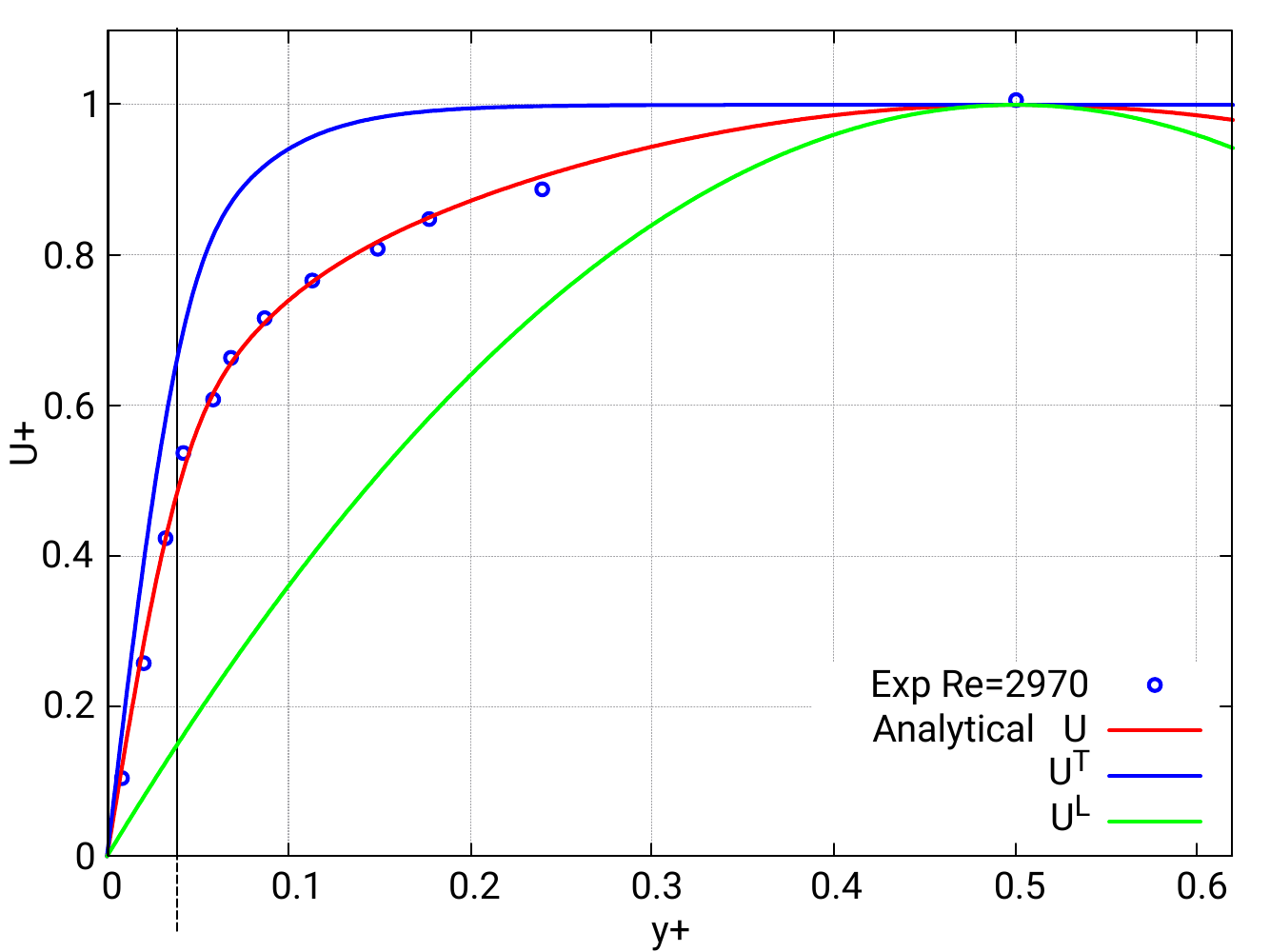}
%\hspace{4cm} (a) \hspace{8cm}(b)
\end{center}
\caption{\label{fig:re2970}Comparison with Wei (1989) experiments at Re = 2970: the mean streamwise velocity  (circles), the Navier-Stokes laminar solution $U^L$ for the streamwise velocity (parabolic function, green line),  the turbulent $U^T$ solution the streamwise velocity (super-exponential function, blue line), 
and their superposition $U$, the solution of the Alexeev hydrodynamic equations. The solution $U$ fits the experimental data well. The vertical black line shows the value of $\delta$.}
\end{figure}
%
%%%%%%%%%%%%%%%%%%%%%%%%%%%%%%%%%%%%%%%%%%%%%%%%%%%%%%%%%%%%%%%%%%%%%%%%%%%%%%%%%%
%
%    Section Comparison of analytical solutions with experiments
%
%%%%%%%%%%%%%%%%%%%%%%%%%%%%%%%%%%%%%%%%%%%%%%%%%%%%%%%%%%%%%%%%%%%%%%%%%%%%%%%%%%

\section{Comparison of analytical solutions with experiments \label{sec:comparison}}
%%%%%%%%%%%%%%%%%%%%%%%%%%%%%%%%%%%%%%%%%%%%%%%%%%%%%%%%%%%%%%%%%%%%%%%%%%%%%%%%%%
%
%    Subsection comparion for channel flows
%
%%%%%%%%%%%%%%%%%%%%%%%%%%%%%%%%%%%%%%%%%%%%%%%%%%%%%%%%%%%%%%%%%%%%%%%%%%%%%%%%%%
\subsection{Comparison of solutions for channel flow: Wei and Willmarth  experiments \label{sec:comp_channel_wei}}
%===========================================================================

We use experimental data for turbulent flow in a two-dimensional channel to compare solutions obtained from the Navier-Stokes equations and the Alexeev Hydrodynamic Equations. Only stationary solutions for incompressible flow are considered, under the assumption that these yield the mean flow velocity. The solutions are compared with high-quality experimental data from multiple sources.

%%%%%%%%%%%%%%%%%%%%%%%%%%%%%%%%%%%%%%%%%%%%%%%%%%%%%%%%%%%%%%%%%%%%%%%%%%
%
%       Subsubection Wei and Willmarth experiment
%
%\subsubsection{Wei and Willmarth experiment}
The experiments of Wei and Willmarth (1989) \cite{Wei_1989} were done in turbulent channel flow over the range of four Reynolds number from 2970
to 39582 based on a channel half-width.  The working fluid was distilled water. The analytical solutions and the experimental data for streamwise mean velocity at Reynolds number $Re=2970$ are shown in Figure \ref{fig:re2970}. The experimental velocity data are shown by circles. 

Figure shows both  the $U^L$, parabolic solution (laminar, green line), that is the Navier-Stokes solution, and $U^T$, super-exponential (turbulent solution, blue line) components, and a solution of Alexeev hydrodynamics equations,
$U$ (red line), that is a linear superposition of $U^L$ and $U^T$ for $\gamma=0.65$  and $\delta=0.04$. The solution of Alexeev hydrodynamics equations provides better fit to experimental data, than the Navier-Stokes solution.

The vertical black line shows the value of $\delta$, that presents the boundary layer thickness scale, where the linear velocity profile ends. It was found \cite{Fedoseyev_1998a, Fedoseyev_1998b} that  $\delta$  coincides with the Kolmogorov microscale, observed in experiment by Koseff and Street (1984) for 3D lid driven cavity flow \cite{Koseff_1984}.
 
\begin{figure}
\begin{center}
\includegraphics[width=0.65\textwidth]{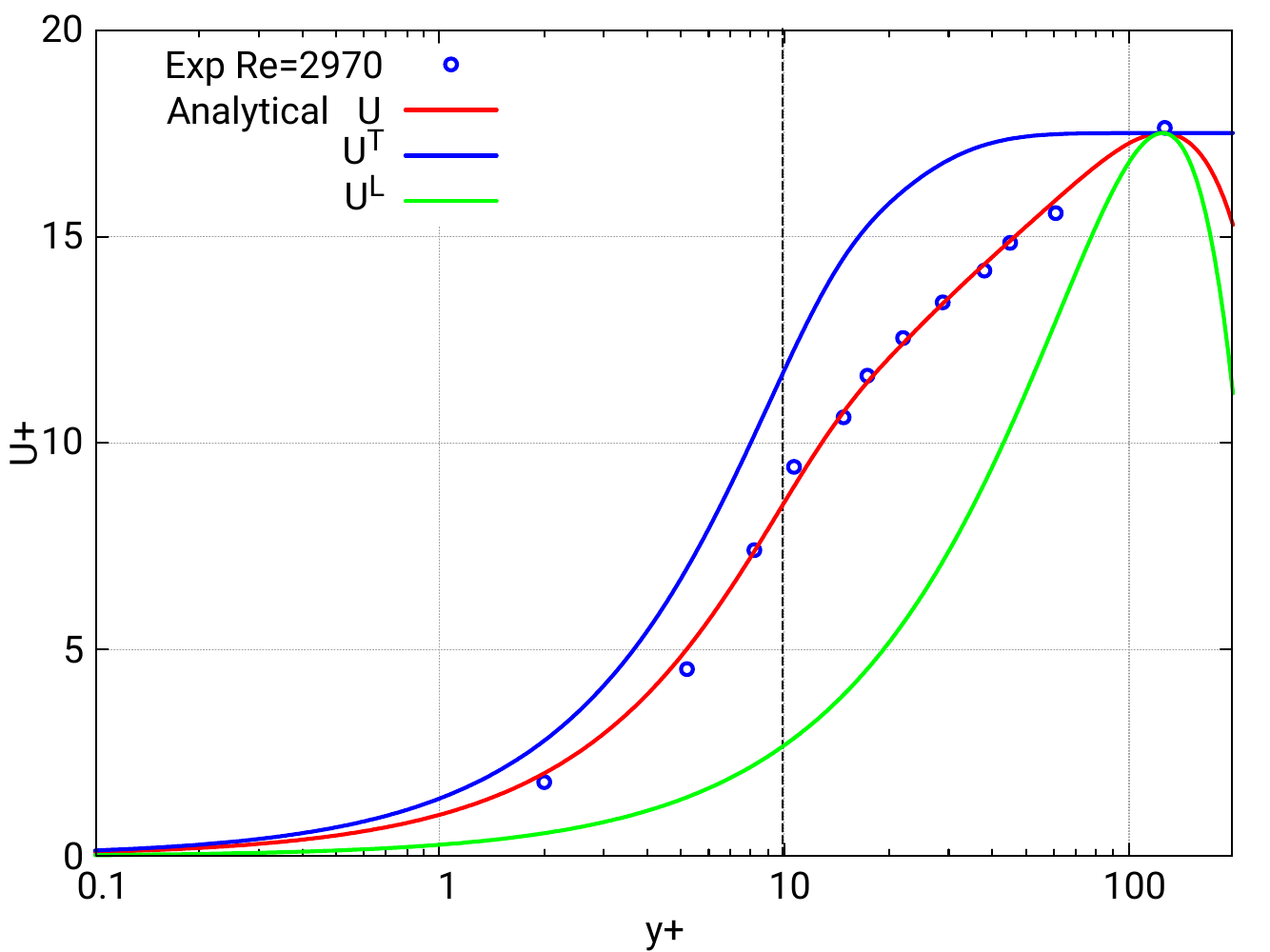}
\end{center}
\caption{\label{fig:re2970log}Comparison of streamwise mean velocity from  Wei (1989) experiments in (Y+, U+) coordinates at Re = 2970 (circles) with the Navier-Stokes laminar solution $U^L$ (parabolic solution, green line),  THE turbulent solution $U^T$ (super-exponential solution, blue line), 
and their superposition $U$ (solution of the Alexeev hydrodynamic equations). The vertical dashed black line indicates $\delta$.}
\end{figure}

\begin{figure} 
\begin{center}
\includegraphics[width=0.65\textwidth]{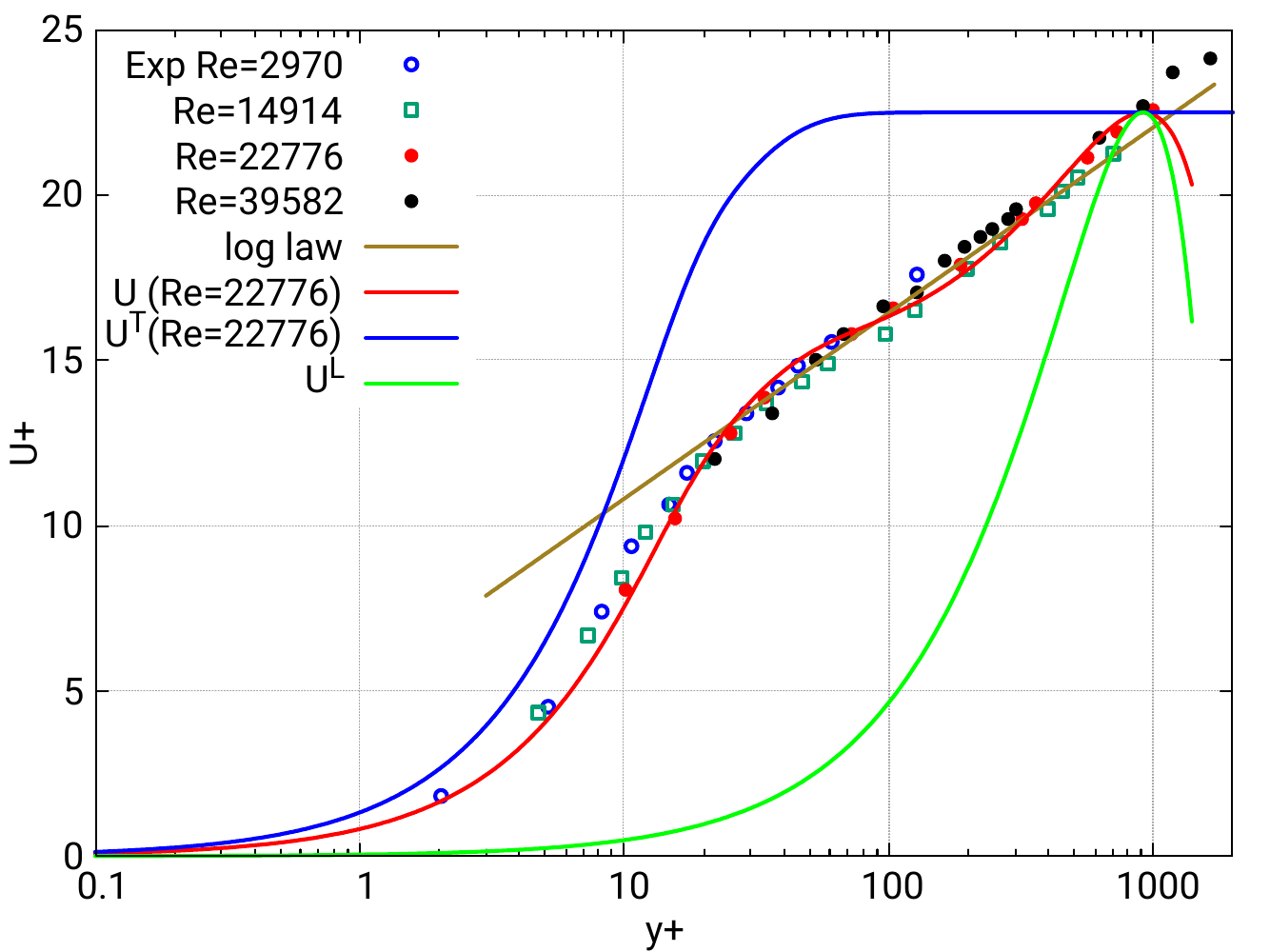}
\end{center}
\caption{
 Velocity profiles for turbulent flow in channel: the experimental data (dots) by Wei (1989) \cite{Wei_1989}, analytical solution of the Navier-Stokes laminar solution (parabolic function, green line) $U^L$, turbulent $U^T$ solution (super-exponential function, blue line), and Alexeev hydrodynamic equations solution $U$ (red line), the superposition of $U^L$ and $U^T$, for Reynolds number Re=22776, that fits well the experimental data (red dots) for Re=22776. The log law by von Karman, $U^+ = 1/k~log (y^+) + B$ (log law) is  shown as olive-green line.}
\label{fig:wei_all}
\end{figure}

We also plot the velocities in $(U^+, y^+)$ coordinates, as presented in Wei (1989) paper.
The parameter $y^{+}= {y U_{\tau}}/{\nu}$, where $U_{\tau}$ is so called
friction velocity, y is the absolute distance from the wall, and $\nu$
is the kinematic viscosity. One can interpret $y^+$ as a local Reynolds
number. The friction velocity $U_{\tau}$ is defined as 
\begin{equation}\label{eq:utau}
u_{\tau}=\sqrt{ {\tau_{w}}/{\rho}},
\end{equation}

\noindent where wall shear stress $\tau_w$, $\tau_{w}=\rho\nu\frac{dU}{dy}$
at y=0, and the dimensionless velocity is given by $U^{+}= {u}/{U_{\tau}}$.
The Figure \ref{fig:re2970log} demonstrates that 
the superposition $U$ provides an excellent fit to the experimental mean velocity profile in $(U^+, y^+)$ as well. The vertical dashed black line shows the value of $\delta$.
Similar results for Re=22776 are presented in Figure \ref{fig:wei_all}, where the Navier-Stokes solution is far from the experimental data, and the solution of the Alexeev hydrodynamic equations is much closer to the data. Such $(U^+, y^+)$ plots often provide the comparison with the classical logarithmic von Karman law,
\begin{equation}\label{eq:loglaw}
U^+ = \frac{1}{k}~log (y^+) + B, 
\end{equation}
with k = 0.41, B = 5.2 (sometimes with slight variations).

The comparison with Wei (1989) experiments at other Reynolds numbers, and with another channel experiment by Pasch (2023) \cite{Pasch_2024} are provided in \cite{Fedoseyev_2025}. 
% \cite{Fedoseyev_2025}. 
%================================================================================. 

%%%%%%%%%%%%%%%%%%%%%%%%%%%%%%%%%%%%%%%%%%%%%%%%%%%%%%%%%%%%%%%%%%%%%%%%%%%%%%%%%%
%
%       Subsection Van Doorne pipe experiment
%
\subsection {Comparison of solutions for pipe flow}
\subsubsection{Van Doorne and Westerweel pipe experiment.}

\begin{figure}[ht!]
\begin{center}
\includegraphics[width=0.60\textwidth]{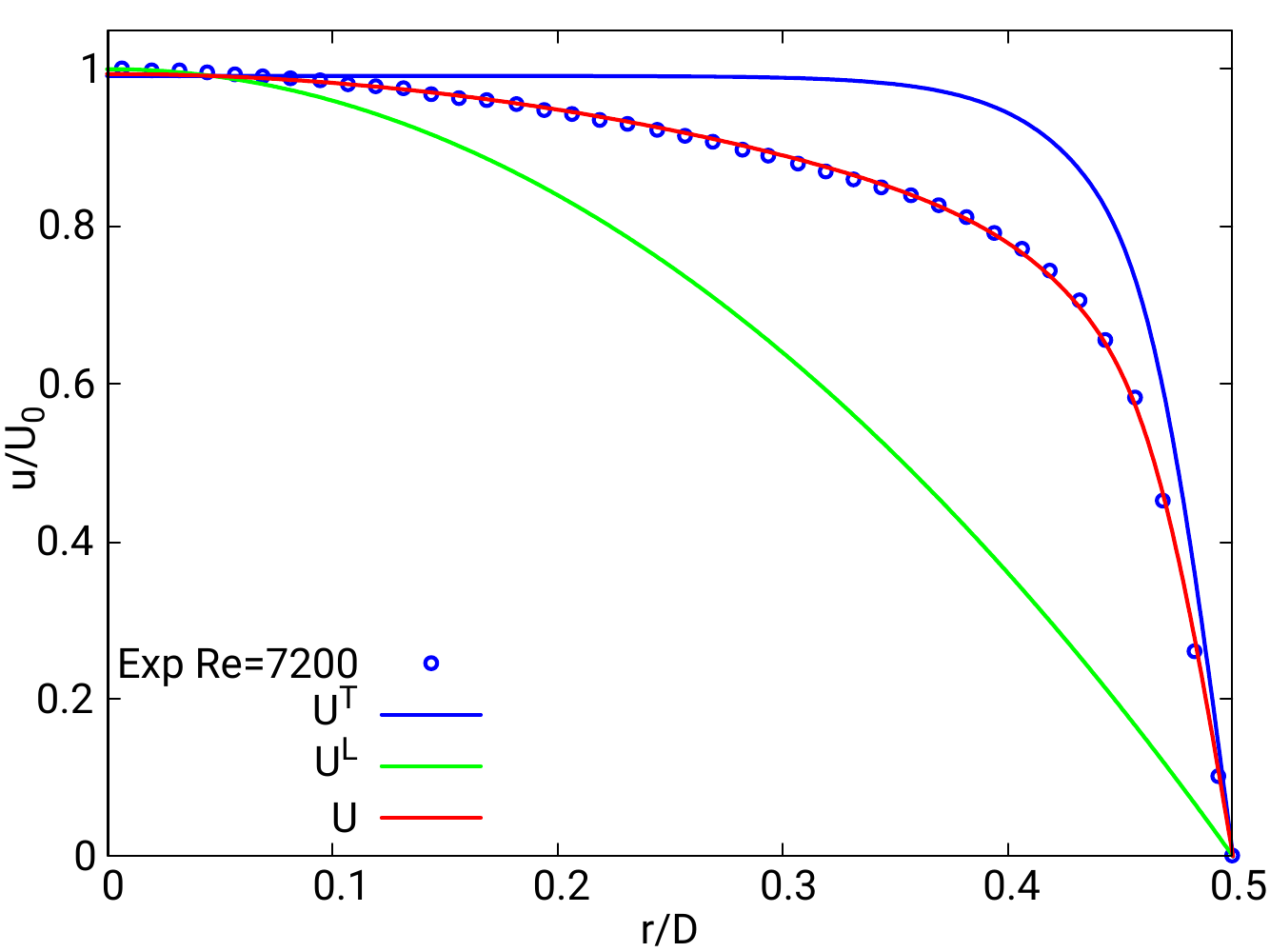}
\end{center}
\caption{\label{fig:doorne}
 Comparison of experimental data for streamwise mean velocity $U_{exp}=u/U_0$ versus radius $r/D$  ($D$ is the diameter)  (blue circles) for the van Doorne experiment \cite{Doorne_2007} at Re=7200 with the laminar (parabolic) solution $U^L$ of the Navier-Stokes equations (green line), the turbulent (super-exponential) solution $U^T$ (blue line), and the solution $U$ of the Alexeev hydrodynamic equations (red line).  The  analytical solution $U$ fits the experimental velocity profile well, while other solutions deviate significantly from the data. }
\end{figure}

In van Doorne and Westerweel (2007) experiment, a circular pipe with an inner diameter of 40 mm and a total length of 28 m was used, so the ratio of diameter to length was $D/Length=700$ \cite{Doorne_2007}. The working fluid was tap water. Due to a well designed contraction and thermal isolation of the pipe, the flow can be kept laminar up to Re = 60,000. It was found that the Coriolis force affects the velocity profile, as leads to an asymmetry for Re > 3,000, as explained by Draad (1998) \cite{Draad_1998}, so the orientation of the  experimental pipe was chosen to exclude the Coriolis force. The measurements at Re=7200 were carried out at the distance of 26 m from the inlet, and stereoscopic-PIV was used \cite{Doorne_2007}.  The analytical solution (\ref{eq:AHEsol}) can be applied to circular pipe flow, when $\delta= \sqrt{\tau_0 \nu} / L_0 \ll 1$, which is the case, as $\delta=0.047$. Figure \ref{fig:doorne} shows the experimental data for the streamwise mean velocity digitized from \cite{Doorne_2007} along with several plots: (i) Navier-Stokes solution, the laminar (parabolic) flow profile (green line), (ii) turbulent (superexponential) flow profile (blue line) 
and (iii) the AHE solution (red line). The value of $\gamma$ is  $\gamma=0.68$, obtained from minimal viscous dissipation principle \cite{Fedoseyev_2024}. One can see that neither the laminar nor turbulent solution fit the
data, but the superposition given by equation (\ref{eq:AHEsol}), the AHE solution (red line), provides good comparison to the experimental data.

%%%%%%%%%%%%%%%%%%%%%%%%%%%%%%%%%%%%%%%%%%%%%%%%%%%%%%%%%%%%%%%%%%%%%%%%%%%%%%%%%%
%
%       Subsubection Zagarola pipe experiments
%
%%%%%%%%%%%%%%%%%%%%%%%%%%%%%%%%%%%%%%%%%%%%%%%%%%%%%%%%%%%%%%%%%%%%%%%%%%%%%%%%%%
\subsubsection{Superpipe experiments by Zagarola, Smits, Orszag and Yakhot.}
A series of 26 experiments was conducted using the Princeton Superpipe setup, with air as the working fluid \cite{Zagarola_1996}. The circular pipe diameter was 12.9 cm.  The temperature was near ambient (295-300 K), and the pressure was varied between 1 and 186 atmospheres for 26 different experiments at 
Reynolds numbers, ranging from $3.15 \cdot 10^4$ to $35.2 \cdot 10^6$. The effects of the Coriolis force which potentially can affect the results, according to \cite{Doorne_2007} and \cite{Draad_1998}, were not discussed by the authors of \cite{Zagarola_1996}.

The experimental conditions (pressure, density, viscosity, pressure gradient and timescale $\tau$) varied by several orders of magnitude. However, the analytical solution parameters, $\gamma$ and $\delta$, did not vary significantly and remained within $0.6 < \gamma <0.7$ and $0.003 <  \delta < 0.006$.

Analytical solutions were constructed for all 26 experiments, and three representative examples are presented for comparison at the lowest, highest, and intermediate Reynolds numbers.

\begin{figure}[ht!]
\begin{center}
\includegraphics[width=0.60\textwidth]{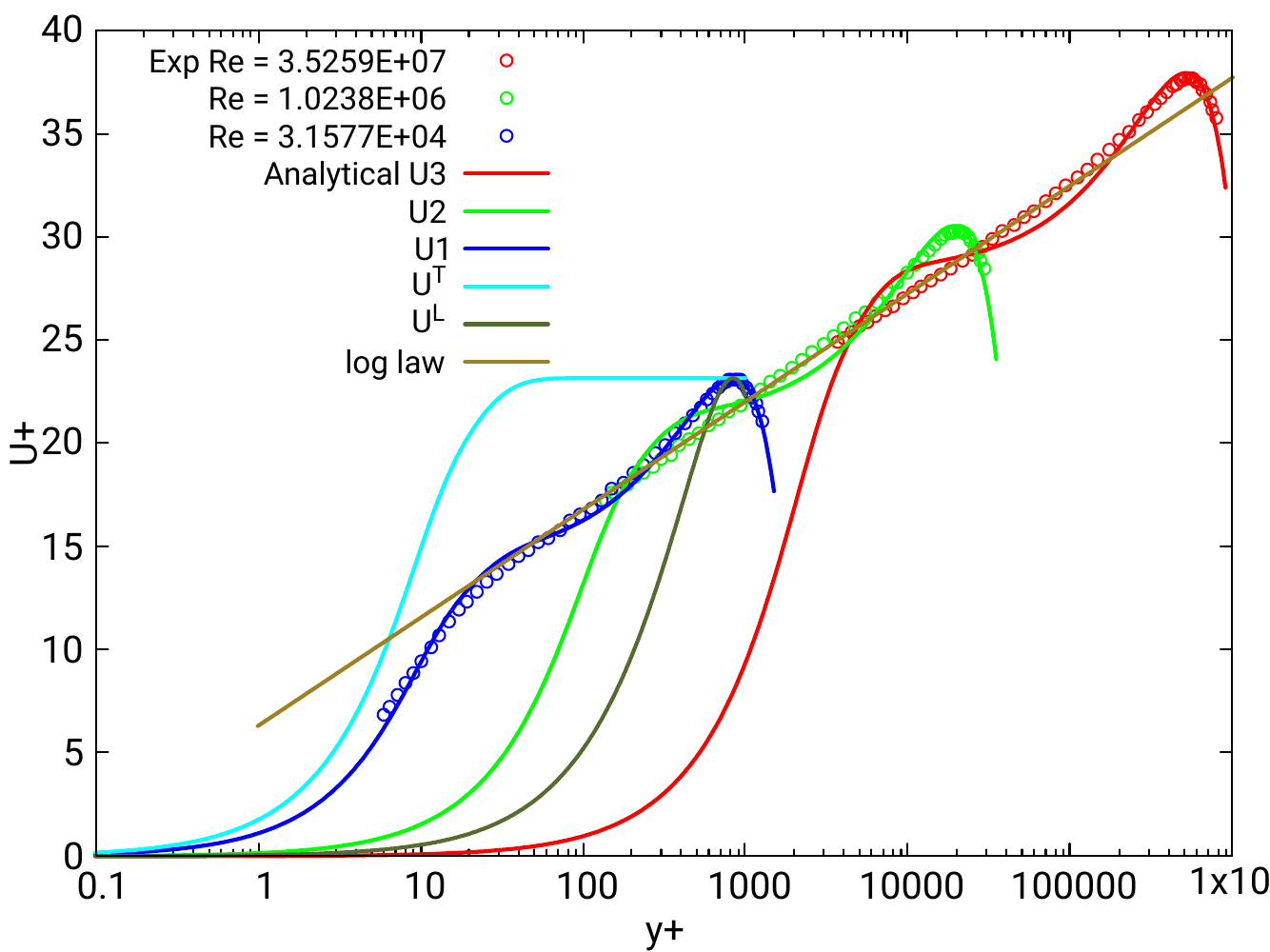}
\end{center}
%\vspace{-7mm}
\caption{\label{fig:super}
Comparison of the streamwise velocity in turbulent pipe flow from the Princeton Superpipe experiment \cite{Zagarola_1996}, $Re=3.15\cdot 10^4$ (blue circles), $Re=1.02\cdot 10^6$ (green circles), and $Re=3.52\cdot10^7$ (red circles). The  analytical solution $U$ is shown for each case in the corresponding color. The laminar solution $U^L$  and turbulent solution $U^T$ are shown for $Re=3.15\cdot 10^4$ only. The log law by von Karman, $U^+ = 1/k~log (y^+) + B$ (log law) is shown by olive-green line. The maximum error of  3\% for AHE solution was observed at the highest Reynolds number.}
\end{figure}

Figure \ref{fig:super} shows the streamwise velocity in turbulent Superpipe experiment for the minimum Reynolds number $Re=3.15\cdot 10^4$, the intermediate $Re=1.02\cdot 10^6$, and the maximum $Re=35.2\cdot10^6$, along with the analytical solution $U$ for each case. The laminar solution $U^L$ (Navier-Stokes) and turbulent solution $U^T$ are shown only for the experiment at $Re=3.15\cdot 10^4$. 

The von Karman log law is also shown (olive - green line).  The coefficients for the von Karman law (\ref{eq:loglaw}) differ here: k=0.44 and B=6.3, as chosen by the  authors of the experiments \cite{Zagarola_1996}. The von Karman law is limited to the region where the fluid mixes intensively, missing the near-wall and near-middle (buffer layer) region at $y^+ < 30$ and the outer region (nonlinear and essentially inviscid) that  begins at $y^+ > 200$ (for $Re=3.15\cdot 10^4$). The log-law error in the buffer layer is 30\%, and the error in the outer region is 6\%.

The AHE analytical solution shows agreement with the experimental data within 3\% as observed at the highest Reynolds number, while the other solutions ($U^L$ and $U^T$) deviate significantly from the data.

%%%%%%%%%%%%%%%%%%%%%%%%%%%%%%%%%%%%%%%%%%%%%%%%%%%%%%%%%%%%%%%%%%%%%%%%%%%%%%%%%%
%
%    Section Discussion
%
%%%%%%%%%%%%%%%%%%%%%%%%%%%%%%%%%%%%%%%%%%%%%%%%%%%%%%%%%%%%%%%%%%%%%%%%%%%%%%%%%%
\section{Discussion\label{sec:discussion}}
%%%%%%%%%%%%%%%%%%%%%%%%%%%%%%%%%%%%%%%%%%%%%%%%%%%%%%%%%%%%%%%%%%%%%%%%%%%%%%%%%%
%
%    Subsection Navier-Stokes solutions
%
%%%%%%%%%%%%%%%%%%%%%%%%%%%%%%%%%%%%%%%%%%%%%%%%%%%%%%%%%%%%%%%%%%%%%%%%%%%%%%%%%%
\subsection{From Boltzmann to Navier-Stokes equations\label{sec::Boltzmann}}

The theoretical models for hydrodynamic flows fall under Hilbert's 6th problem \cite{Hilbert_1902}, addressing the transition from the "atomistic view to the laws of motion of continua". It includes at least two steps: \\
(1) from mechanics to kinetics (from Newton to Boltzmann),\\
 and \\
(2) from kinetics to mechanics and nonequilibrium thermodynamics of continua (from Boltzmann to Euler and Navier-Stokes), Gorban and Karlin (2014) \cite{Gorban_2014}.  

It is considered that the first step, from mechanics to kinetics (from Newton to Boltzmann), was completed with the complexity approach to randomness invented by Solomonoff and Kolmogorov, as remarked in \cite{Gorban_2014} (see the review \cite{Zvonkin_1970} and the textbook \cite{Li_1997}).

As for the second step, from kinetics to hydrodynamics and nonequilibrium thermodynamics of continua (from Boltzmann to Euler and Navier-Stokes), a review of mathematical works in this area shows that the well-known Euler and Navier-Stokes equations are valid only at "a limit of very slow flows with very small gradients of all fields, i.e., almost no flow at all", Gorban and Karlin (2014) \cite{Gorban_2014}.

This may explain why, in the turbulent flow experiments presented above, the NS solutions deviate significantly from the experimental data, as the flow velocities are not small.

%%%%%%%%%%%%%%%%%%%%%%%%%%%%%%%%%%%%%%%%%%%%%%%%%%%%%%%%%%%%%%%%%%%%%
%
%       Subsection superposition coefficient
%
%%%%%%%%%%%%%%%%%%%%%%%%%%%%%%%%%%%%%%%%%%%%%%%%%%%%%%%%%%%%%%%%%%%%%
\subsection{Superposition coefficients $\gamma$}

 The superposition coefficients $\gamma$ and $(1-\gamma)$ were calculated using the  principle of minimal  viscous dissipation of kinetic energy $\varepsilon_{T}$
relative to the shear stress (derivative of $U$) on the wall  \cite{Fedoseyev_2024}. 
The minimum of $\varepsilon_{T}$ gives a solution for $\gamma$: 
\begin{equation}\label{eq:diss_tot}
\varepsilon_{T} =  \frac{ 1}{U_{y}(0)^2} \int^L_0 U_y^2 dy \cdot
\frac{ 1}{L} \int^L_0 U\ dy,
\end{equation}
where $U(y)$ is presented by Eq.(\ref{eq:AHEsol}). 
The value of $\gamma$ obtained by this procedure yields the best agreement between $U(y)$ and the experimental measurements.
 The parameter $\gamma$ was within the interval from 0.6 to 0.7 for most of the cases considered, with both water and air as a working fluid.
%%%%%%%%%%%%%%%%%%%%%%%%%%%%%%%%%%%%%%%%%%%%%%%%%%%%%%%%%%%%%%%%%%%%%
%
%       Subsection The boundary scale length $\delta$
%
%%%%%%%%%%%%%%%%%%%%%%%%%%%%%%%%%%%%%%%%%%%%%%%%%%%%%%%%%%%%%%%%%%%%%
\subsection{Boundary layer thickness scale $\delta$}
The boundary layer thickness scale and the similarity parameter $\delta=\sqrt{\tau_0 \nu}/L_0$ is the most important parameter for the turbulent flows, discovered in the analytical solution.  The value of $\delta$  coincides with the Kolmogorov microscale, observed in experiment by Koseff and Street (1984) for 3D lid driven cavity flow \cite{Koseff_1984}.

The parameter  $\delta$ is a material property and does not depend on the Reynolds number, as shown by the data from the Wei and Willmarth experiment \cite{Wei_1989}: the experimental data for four different Reynolds numbers ranging from Re=2910 to Re=39582 collapse closely onto the same curve, as shown in Figure \ref{fig:wei_all}. 

The value of $\delta$ is approximately 0.6 mm for distilled water and  approximately the same for  air over a wide pressure range. Knowing $\delta$, one can determine the timescale parameter $\tau_0$, which is also a material property (though this parameter to be of lesser importance).
%%%%%%%%%%%%%%%%%%%%%%%%%%%%%%%%%%%%%%%%%%%%%%%%%%%%%%%%%%%%%%%%%%%%%
%
%       Subsection The turbulent boundary layer
%
%%%%%%%%%%%%%%%%%%%%%%%%%%%%%%%%%%%%%%%%%%%%%%%%%%%%%%%%%%%%%%%%%%%%%
\subsection{Turbulent boundary layer}

The analytical solution $U$ from equation (\ref{eq:AHEsol}) represents the turbulent boundary layer across all regions with great accuracy, as shown in Figure \ref{fig:wei_all} (the analytical solution  at Re=22776 is the red line; the Wei (1989) experiment is shown as red dots), with the linear law at the wall in the range $0 < y^+ < 5$, where  $U^L$ is negligible.

In the near-wall  (buffer layer) region at $5 < y^+ < 30$, the solution $U$ is strictly nonlinear, yet the analytical solution fits the flow data well.
In the far-middle (inner) boundary layer region at $30 < y^+ < 200$, $U^{T}$ is almost constant, while the solution (red line) exhibits variation due to the increasing contribution of the laminar component $U^{L}$. In this region, the AHE solution is a parabolic function  that provides a closer fit to the experimental data (red dots) than the classical logarithmic von Karman law (olive-green line).

The outer region (nonlinear and essentially inviscid) begins at $y^+ > 200$ and extends to the centerline, where the analytical solution also shows good agreement with the experimental data.
%%%%%%%%%%%%%%%%%%%%%%%%%%%%%%%%%%%%%%%%%%%%%%%%%%%%%%%%%%%%%%%%%%%%%
%
%       Subsection The source of turbulence
%
%%%%%%%%%%%%%%%%%%%%%%%%%%%%%%%%%%%%%%%%%%%%%%%%%%%%%%%%%%%%%%%%%%%%%
\subsection{Turbulence source and control}
The turbulent component of the streamwise velocity $U^T$ becomes zero, when the transverse velocity is zero in equation (\ref{eq:2d_mom_UT}). In this case, the flow remains laminar and free of turbulence. The transverse velocity $V^T$ is proportional to the transverse pressure gradient $p^T_y$ in equation (\ref{eq:pTsol}). Thus,  the transverse pressure gradient and transverse velocity are the sources of turbulence. Experiments by van Doorne (2007) showed that the flow can be kept laminar up to Re = 60,000 by suppressing  pressure disturbances at the inlet.

Turbulent flow can be converted to laminar flow by eliminating the transverse velocity component. This reveals a practical control mechanism: by applying controlled suction or injection through an array of small holes in the wall, the transverse velocity can be driven toward zero, thereby suppressing turbulence \cite{Fedoseyev_2026}. Suction removes fluid normal to the wall, while injection introduces fluid; both serve to manipulate the transverse flow component.

While the effectiveness of wall suction and injection for turbulence control has been observed experimentally, we provide the first rigorous theoretical explanation of the underlying mechanism, connecting transverse velocity suppression directly to the transition from turbulent to laminar flow. Remarkably, the suction or injection velocities required are approximately one thousand to ten thousand times smaller than the main streamwise velocities \cite{Fedoseyev_2026, Schlichting_1960}.

%%%%%%%%%%%%%%%%%%%%%%%%%%%%%%%%%%%%%%%%%%%%%%%%%%%%%%%%%%%%%%%%%%%%%%%%%%%%%%%%%%
%
%    Section Conclusion
%
%%%%%%%%%%%%%%%%%%%%%%%%%%%%%%%%%%%%%%%%%%%%%%%%%%%%%%%%%%%%%%%%%%%%%%%%%%%%%%%%%%
%
\section{Conclusion\label{sec:conclusion}}
The analytical solution of the Alexeev Hydrodynamic Equations for turbulent channel flow demonstrates significantly better agreement with the mean velocity experimental data than solutions obtained from the Navier-Stokes equations. These solutions successfully capture the correct velocity behavior across the entire turbulent boundary layer and into the external flow, spanning from the inner viscous sublayer to the outer layer of the boundary layer. The analytical solution reveals an important boundary layer thickness scale through the similarity parameter $\delta$, which coincides with the Kolmogorov microscale observed in experiments. The mechanisms for turbulence generation were identified and a turbulence control method was proposed.
%
%       Section bibliography
%
%\reftitle{References}

%%%%%%%%%%%%%%%%%%%%%%%%%%%%%%%%%%%%%%%%%%
%%%\PublishersNote{}
%\isPreprints{}{% This command is only used for ``preprints''.
%%%\end{adjustwidth}
%} % If the paper is ``preprints'', please uncomment this parenthesis.

\end{document}